\pdfoutput=1
\documentclass[a4paper]{spie}  

\usepackage{amsmath,amsfonts,amssymb}
\usepackage{graphicx}
\usepackage[colorlinks=true, allcolors=blue]{hyperref}

\title{LBT SOUL data as a science test bench for MICADO PSF-R tool}

\author[a,*]{Matteo Simioni}
\author[a]{Carmelo Arcidiacono}
\author[b]{Roland Wagner}
\author[a]{Andrea Grazian}
\author[a]{Marco Gullieuszik}
\author[c]{Elisa Portaluri}
\author[a]{Benedetta Vulcani}
\author[a]{Anita Zanella}
\author[d]{Guido Agapito}
\author[e]{Richard Davies}
\author[f]{Tapio Helin}
\author[g]{Fernando Pedichini}
\author[g]{Roberto Piazzesi}
\author[d]{Enrico Pinna}
\author[b,h]{Ronny Ramlau}
\author[d]{Fabio Rossi}
\author[f]{Aleksi Salo}
\affil[a]{INAF - Osservatorio Astronomico di Padova, Vicolo dell'Osservatorio 5, Padova, Italy, I-35122} 
\affil[b]{Industrial Mathematics Institute, Johannes Kepler University Linz, Altenberger Strasse 69, Linz, Austria, 4040} 
\affil[c]{INAF - Osservatorio Astronomico d'Abruzzo, Via Mentore Maggini, Teramo, Italy, I-64100} 
\affil[d]{INAF - Osservatorio Astrofisico di Arcetri, Via E. Fermi 5, Firenze, Italy, I-50125} 
\affil[e]{MPE - Max-Planck-Institut für extraterrestrische Physik Giessenbachstrasse 1, Garching, Germany, D-85748} 
\affil[f]{LUT University, P.O.Box 20, Lappeenranta, Finland, FI-53851} 
\affil[g]{INAF - Osservatorio Astronomico di Roma, Via Frascati 33, Monte Porzio Catone, Italy, I-00078} 
\affil[h]{RICAM - Johann Radon Institute for Computational and Applied Mathematics, Altenberger Strasse 69, Linz, Austria, 4040} 

\authorinfo{Further author information: (Send correspondence to M.Simioni)\\M.Simioni: E-mail: matteo.simioni@inaf.it}

\begin{document} 
\maketitle

\begin{abstract}
Current state-of-the-art adaptive optics (AO) provides ground-based, diffraction-limited observations with high Strehl ratios (SR). However, a detailed knowledge of the point spread function (PSF) is required to fully exploit the scientific potential of these data. This is even more crucial for the next generation AO instruments that will equip 30-meter class telescopes, as the characterization of the PSF will be mandatory to fulfill the planned scientific requirements. For this reason, there is a growing interest in developing tools that accurately reconstruct the observed PSF of AO systems, the so-called PSF reconstruction. In this context, a PSF-R service is a planned deliverable for the MICADO@ELT instrument and our group is in charge of its development.
In the case of MICADO, a blind PSF-R approach is being pursued to have the widest applicability to science cases. This means that the PSF is reconstructed without extracting information from the science data, relying only on telemetry and calibrations. While our PSF-R algorithm is currently being developed, its implementation is mature enough to test performances with actual observations. In this presentation we will discuss  the reliability of our reconstructed PSFs and the uncertainties introduced in the measurements of  scientific quantities for bright, on-axis observations taken with the SOUL+LUCI instrument of the LBT. This is the first application of our algorithm to real data. It demonstrates its readiness level and paves the way to further testing.
Our PSF-R algorithm is able to reconstruct the SR and full-width at half maximum of the observed PSFs with errors smaller than $2\%$ and $4.5\%$, respectively. We carried out the scientific evaluation of the obtained reconstructed PSFs thanks to a dedicated set of simulated observations of an ideal science case.
  
\end{abstract}

\keywords{Adaptive optics, Astronomy, Infrared imaging, Optical transfer functions, Point spread functions}

\section{INTRODUCTION}\label{sec:intro}  
The exquisite imaging capabilities, both in the visible and near-infrared, expected for the next generation instruments that will equip ground-based astronomical facilities will rely extensively in Adaptive Optics (AO)\cite{2000PASP..112..315W, 2012ARA&A..50..305D, 2016SPIE.9908E..1ZD, 2018SPIE10702E..09D, 2020SPIE11448E..2WA, 2020SPIE11447E..1RR}. In fact, the ability to provide high Strehl ratio (SR), almost diffraction-limited observations is a prerogative to match, and even surpass, the performances of present space telescopes. Nonetheless, a detailed knowledge of the point spread function (PSF) for these AO-assisted instruments is usually required in order to fully exploit their scientific potential. In particular they are characterized by a complex PSF with a high number of degrees of freedom, that is not constant in the field of view and in time and that is wavelength-dependent. 
For this reason, there is a growing interest in tools to infer the PSF of AO-assisted instrument\cite{2012ARA&A..50..305D,2016SPIE.9908E..1ZD, 2018SPIE10702E..09D, 2020SPIE11447E..1RR} and different approach has been proposed. They can be grouped into three main categories: 1) those based on focal-plane data; 2) those based on pupil-plane data (e.g. PSF reconstruction -- hereafter PSF-R); 3) hybrid techniques that combine 1) and 2) hence optimizing the reconstruction. We remand the interested reader to the detailed review of Ref.~\citenum{2020SPIE11448E..0AB}.
This is even more true for the instruments of $30$-meter class telescope as the Multi-Adaptive Optics Imaging Camera for Deep Observations (MICADO), the first light instruments of the European Southern Observatory (ESO) Extremely Large Telescope (ELT). In fact, due to the unprecedented dimensions of the ELT primary mirror, the expected PSF in this case will be characterized by an extremely narrow core and a faint halo that will be a factor of $\sim 5$ times larger than that of the PSF of any current AO facility, in the same regime of Strehl ratio. The MICADO PSF will be thus characterized by a prominent first Airy ring but, more importantly, the halos of multiple point sources in the field of view will blend together and form a distributed background. In this specific case, a service that will provide to the final user the observation-specific MICADO reconstructed PSFs is a planned deliverable\cite{2016SPIE.9908E..1ZD}. The algorithm and methods used for this work are, in fact, those developed in this context. A PSF-R approach that allows the reconstruction of the PSF without any needed information coming from the focal plane instrument has been chosen for MICADO\cite{2020SPIE11448E..37S,grazian_psfrstatus}. The PSF is thus reconstructed using instrument telemetry and calibrations, with a wider applicability of the method to different science case. Especially for typical extragalactic targets where few or no useful point sources are present in the science frames for focal-plane methods of PSF estimation. 
The adopted PSF-R method for MICADO is described in Ref.~\citenum{2018JATIS...4d9003W} and is aimed at reconstruct the PSF from the residual phase power spectrum density. Such algorithm has been tested on end-to-end simulations only. 

The present paper focus on the first implementation of the method to real single-conjugated adaptive optics (SCAO) observations. Its performances has been investigated using archival, SCAO data from the Large Binocular Telescope (LBT\cite{lbt}). Specifically, the data consist of two distinct set of observations of bright, on-axis point sources, preliminary described in Ref.~\citenum{2020SPIE11448E..37S}. They both have been taken with the Single conjugated adaptive Optics Upgrade for LBT (SOUL\cite{2016SPIE.9909E..3VP}), using LBT Utility Camera in the Infrared (LUCI\cite{2003SPIE.4841..962S}). The similarities between this instrument and MICADO makes this choice particularly interesting, also in light of the future development of our tool.
To evaluate the accuracy of the reconstructed PSF, we also present an idealized scientific case, showing the gain offered by this method to the photometrical and morphological characterization of a high-$z$ galaxy. The complete analysis is presented in Ref.~\citenum{2022arXiv220901563S}.

The paper is organized as follows: 
the data are presented in Section \ref{sec:data}; information on the PSF-R algorithm are provided in Section \ref{sec:psfr}; in Section \ref{sec:res} the performance of the PSF-R algorithm is discussed evaluating also the impact on the measure of the morphological parameters of an idealized compact galaxy, and Section \ref{sec:fin} is dedicated to the conclusions.

\section{DATA}\label{sec:data}
\begin{table}[htp]
\caption{Log of the SOUL+LUCI@LBT observations} 
\label{tab:log}
\centering    
\begin{tabular}{lcc}
\hline\hline
                          & DAYTIME             & NIGHTTIME          \\
\hline
PROG. ID                  & 1183150             & 1286078            \\
Target                    & illuminated fiber   & HD$\,873$          \\
RA [hh mm ss]             & N.A.                & $00\, 13\, 13.123$ \\
DEC [$^\circ \, ' \, ''$] & N.A.                & $+20\, 45\, 9.479$ \\
R [mag]                   & $10.0$              & $8.57$             \\
Date [yyyy/mm/dd]         & 2019/03/29          & 2019/11/09         \\
Filter                    & H                   & FeII               \\
NDIT$\times$DIT [s]       & $90\times1$         & $20\times0.313$    \\
AIRMASS                   & N.A.                & 1.0248             \\
\hline
\end{tabular}
\end{table}
Archive SCAO LBT data has been used\cite{2021MNRAS.508.1745A} . They consist of 2 independent datasets of near-infrared SOUL+LUCI observations. Additional information on the observations are listed in Table \ref{tab:log}. The first dataset refers to observations of an artificial source and simulated atmospheric condition\cite{2008SPIE.7015E..12R,2010SPIE.7736E..09E} (daytime dataset). A seeing of $1.2$ arcsec was simulated with an average wind speed of $15\,{\rm m/s}$, the resulting saving rate for the associated telemetry is $500\,{\rm frames/s}$. 
The second dataset consists of observations of a natural point source (nighttime dataset). For this dataset, the use of the narrow-band FeII filter, reduces the impact of atmospheric dispersion on data. The associated elongation of the PSF is of the order of few pixels and we take it into account in our PSF-R. The measured seeing was again $1.20$ arcsec, the wind speed $3.4\,{\rm m/s}$. Given the brightness of the target, the frequency of AO correction has an higher value of $1700\,{\rm Hz}$ with respect to the daytime dataset. The associated saving rate for the telemetry is $850\,{\rm frames/s}$, because of the selected decimation (a factor 2).

The observed PSFs for both datasets have been derived from raw data enhancing the final signal-to-noise ratio (SNR) by stacking all the individual science frames. A region of $\sim8\,{\rm arcsec^{2}}$ ($190{\rm px}\times190{\rm px}$), centered at the point source has been considered for each observed PSF, that has been also normalized with respect to its flux. The SNR per pixel for the daytime observed PSF goes from $250$ at peak to a mean value of $10$ at a radial distance of $20{\rm px}$ from PSF centre and reaching mean SNR per pixel$=1$ around $60$px. The nighttime observed PSF has a peak SNR per pixel of $50$, that decreases to a mean value of $10$ at a radial distance of $20{\rm px}$. For this dataset, a mean SNR per pixel$=1$ is reached around $90$px from PSF centre. The different filters used for each dataset and the variation in environmental conditions both contribute to the observed difference in the SNR trends for the two observed PSFs.

\section{PSF RECONSTRUCTION}\label{sec:psfr}

The PSF-R method used to reconstruct the on-axis PSF from AO telemetry data is the one described in Ref.~\citenum{2018JATIS...4d9003W}, which is by itself an update of the classical algorithm developed by Ref.~\citenum{1997JOSAA..14.3057V}.  The high flexibility of our PSF-R  method\cite{2018JATIS...4d9003W} implies that only minor tuning were needed to adapt it to SOUL+LUCI, making it also one of the first, to our knowledge, successful implementation of a PSF-R algorithm to a pyramid wave-front sensor (WFS) AO system. The core idea of this approach is to compute the Optical Transfer Function (OTF), the Fourier transform of the PSF derived from the pupil mask. The advantage of this approach relies in the possibility of factorize the OTF into independent components, wavelength and line-of-sight dependent, which can be either calculated from the wavefront residual phase or using calibrated values. While an extensive mathematical description of the method is present in Ref.~\citenum{2018JATIS...4d9003W}, it is worth noting that in this specific case the assumptions of fast frame rate and a least-squares reconstructor are fulfilled. Therefore, the same decomposition of Ref.~\citenum{1997JOSAA..14.3057V} can be performed on the post-AO OTF, which is related to the effects of the AO corrected atmospheric turbulence.
The time and wavelength dependent PSFs are then obtained transforming the total OTF of the system via inverse Fourier transform. They are finally integrated in wavelength over the pass-band of the selected filter to obtain the reconstructed PSFs. 

Focusing only on on-axis point source, the anisoplanatic component can be neglected. Moreover, since for both dataset the loop was sampled at frequencies $>500\,$Hz, and non-negligible physical vibrations of the instrument occur at much slower timescale, it can be assumed that the SOUL+LUCI system already compensate them.

As anticipated, we only used the telemetry data associated with the scientific observations to reconstruct the PSFs. The telemetry data provided by the SOUL team include: AO WFS slopes data history, control matrix, interaction matrix, gain vector, and the pupil definition. In particular, the pupil shape for daytime is different with respect to that of nighttime due to the different instrument setups. We take it into account when reconstructing the PSFs.
Furthermore, non-common path aberrations between WFS optical train and the LUCI camera are compensated by the AO system making the pyramid loop closing on non-zero reference slopes. The small residuals (depending on the pyramid optical gain \cite{Korkiakoski} and un-calibrated reciprocal flexures between the camera and the WFS) has been measured to have values of about $40\,{\rm nm}$ in both daytime and nighttime. Even if negligible, these estimated wavefront errors were also applied in the PSF-R algorithm.

\section{Discussion}\label{sec:res}
\begin{table}[ht]
\centering 
\caption{Properties of the observed and reconstructed PSFs for both datasets and for both the observed (OBS) and reconstructed (REC) PSF. The list include the Strehl ratio, full-width at half maximum (FWHM) and the encircled energy in the core ($\rm{EE_{core}}$). The radius of the circular aperture, in this case, has been fixed to $2.7\,{\rm px}$: the expected FWHM of the diffraction limited PSF. The radius of the circular aperture, centerd on the PSF, that contains $50\%$ of the total flux ($\rm{R_{50}}$) is also provided. The relative error of the reconstruction is provided for each parameter.}
\label{tab:orprop}
\begin{tabular}{lccc}
\hline\hline
Parameter           & OBS.  & REC.  & DIFF. $[\%]$ \\
\hline
DAYTIME             &       &       &              \\
\hline
SR                  & 0.59  & 0.58  & 1.7          \\
FWHM[mas]           & 44.10 & 43.80 & 0.7          \\
${\rm EE_{core}}$   & 0.43  & 0.43  & 1.9          \\
${\rm R_{50}}$[mas] & 50.23 & 53.97 & 7.4          \\
\hline
NIGHTTIME           &       &       &              \\
\hline
SR                  & 0.58  & 0.59  & 1.7          \\
FWHM[mas]           & 43.80 & 45.75 & 4.4          \\
${\rm EE_{core}}$   & 0.43  & 0.45  & 4.0          \\
${\rm R_{50}}$[mas] & 52.92 & 50.83 & 3.9          \\
\hline
\end{tabular}
\end{table}
\begin{figure*}
  \centering
     \includegraphics[width=0.45\textwidth]{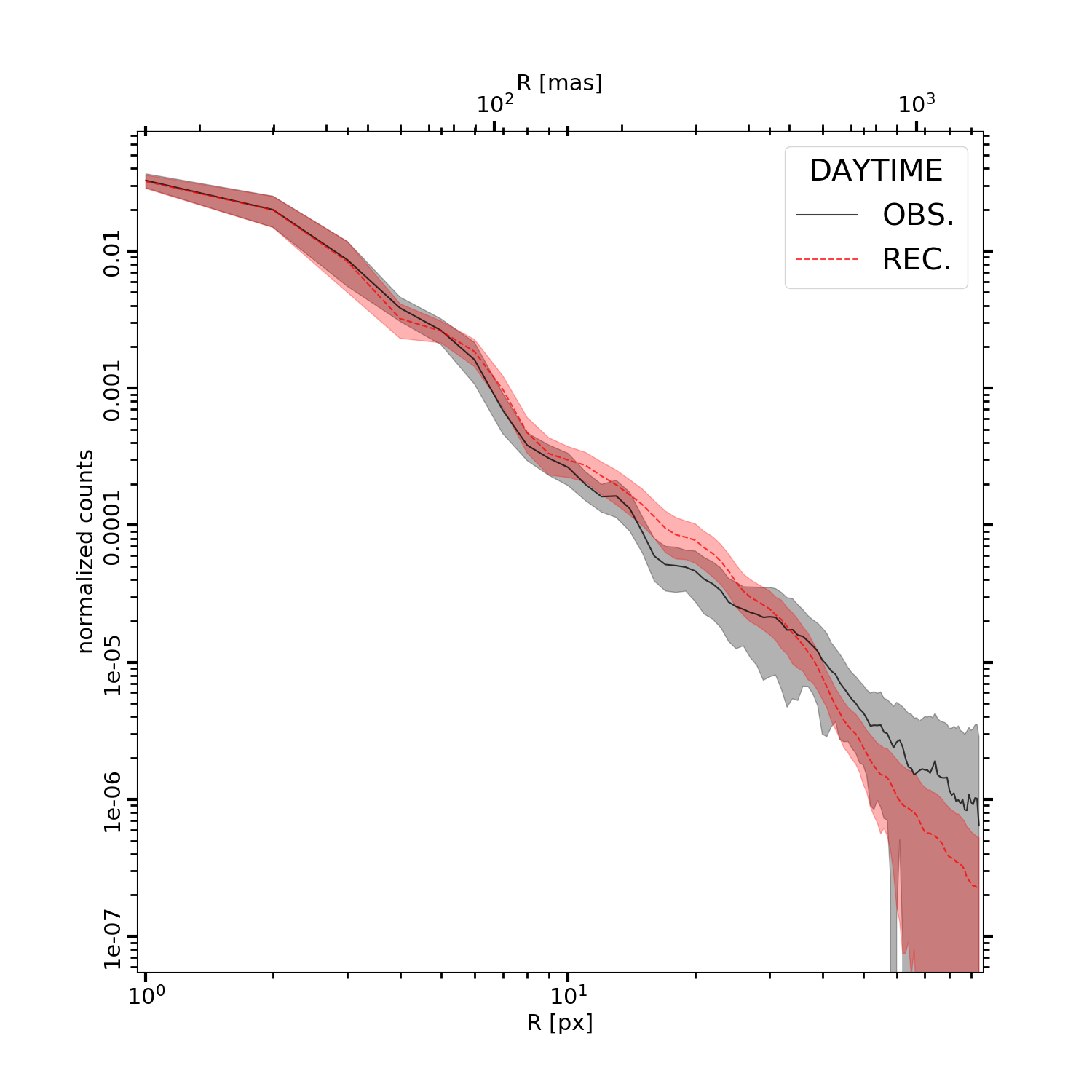}
     \includegraphics[width=0.45\textwidth]{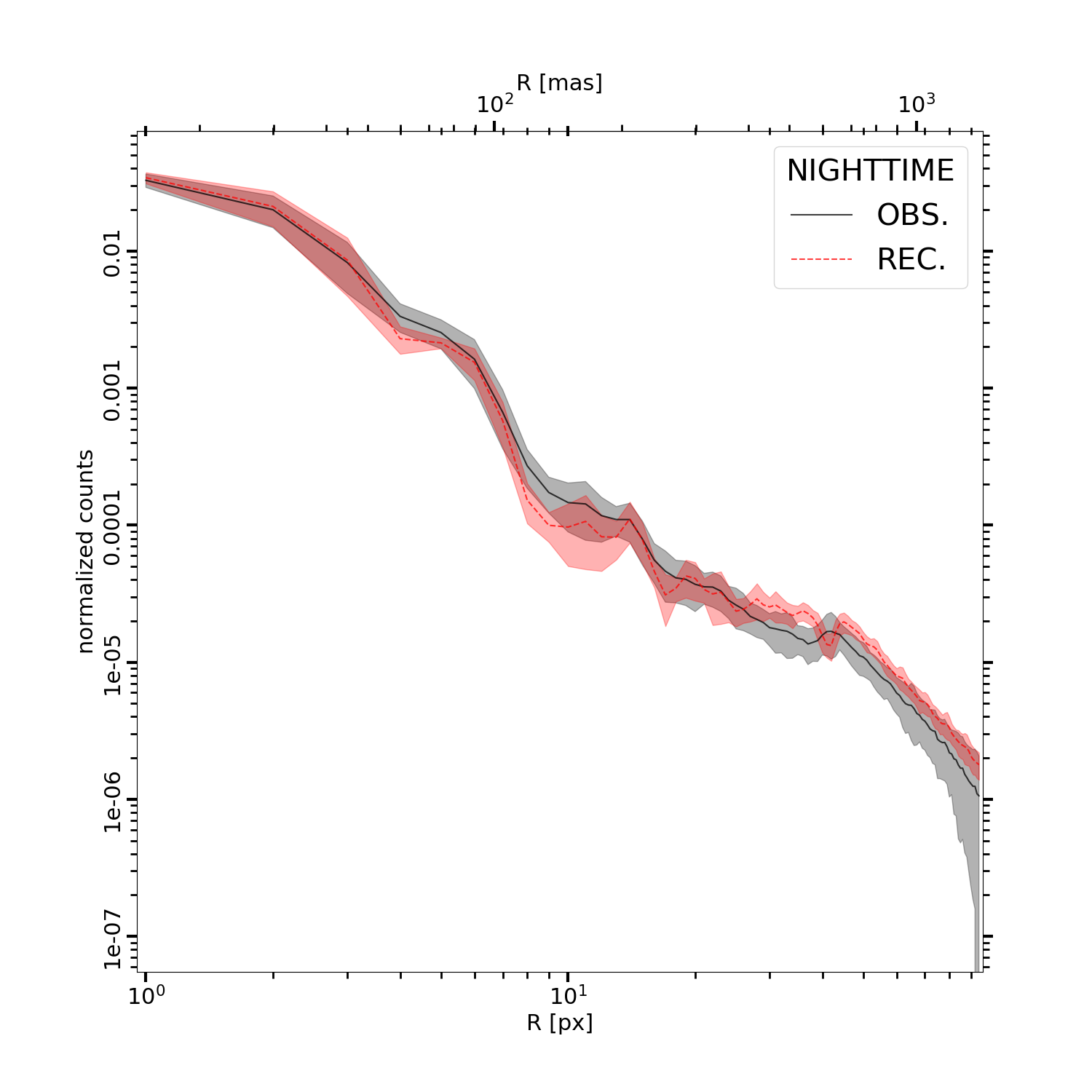}
     \caption{Radial profile (mean over all angles) of the observed PSF (black) and the reconstructed PSF (red). Left panel. Daytime. Right panel. Nighttime. The associated standard deviations are indicated with shaded regions. }
     \label{fig:rpdn}
\end{figure*}
A quantitative analysis of the PSFs, both observed and reconstructed, corroborates the good match between the observed and reconstructed PSF. Table~\ref{tab:orprop} lists a collection of the most relevant PSFs properties. For both datasets, the SR is recovered with an accuracy of $1.7\%$. 
in Figure \ref{fig:rpdn}, we compare the radial profile of the observed and reconstructed PSFs. This allows a further discussion of the results.
For both datasets, the radial profiles of the reconstructed PSFs match the observed one with good accuracy, particularly in the central regions. This is confirmed by the relative errors on the FWHM of the reconstructed PSFs, that are accurate at the level of $1\%$ and $4\%$ for daytime and nighttime, respectively. The absolute difference is of the order of $0.1\,{\rm px}$) in the worst case. The accuracy of the reconstruction is further corroborated by the EE$_{core}$ values, a precision better than $4\%$ is reached in both cases. Moving outwards to around $50\,{\rm mas}$, the light distribution of the reconstructed PSF keeps following the observed one with errors on ${\rm R_{50}}$ of the order of $7\%$ and $4\%$ for daytime and nighttime respectively. Again the absolute difference in ${\rm R_{50}}$ are of the order of $0.2\,$px in the worst case. For larger apertures, small deviations are detected. Notably, the bump in the radial profile of the nighttime reconstructed PSF at a distance of $\sim40\,$px coincides with the expected position of the AO control radius. This feature, however, does not seem to affect significantly the match between observed and reconstructed PSF.

\begin{figure*}[t]
  \centering
     \includegraphics[width=0.9\textwidth]{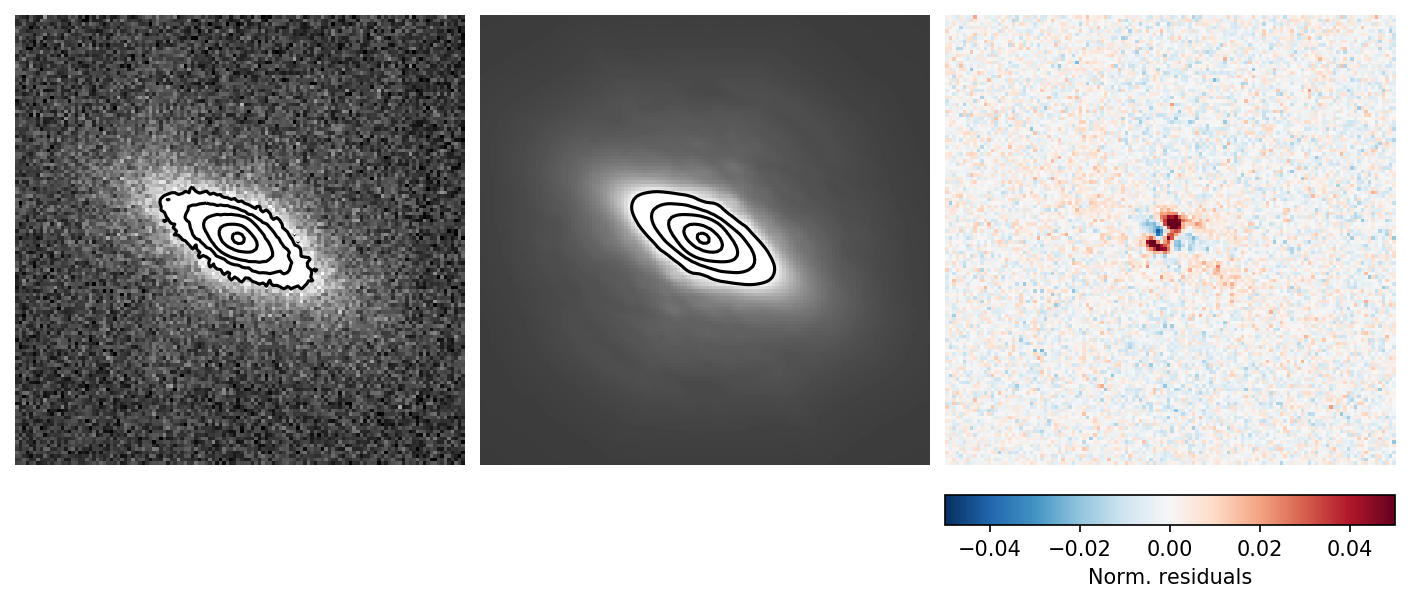} 
     \caption{Simulated high-$z$ galaxy. A $128{\rm px}\times128{\rm px}$ zoom-in is shown. The nighttime observed PSF is used as input to generate the simulated observation (left panel). The best-fit GALFIT S\'ersic model is shown in the central panel. In this case, the associated reconstructed PSF is used as the input PSF for GALFIT. The contours refer to the same count levels as those used in the left panel. Normalized residuals are then shown in the right panel. The colorbar for the normalized residuals is shown on the bottom. }
     \label{fig:nscier}
\end{figure*}
\begin{table*}[ht]
\centering
\caption{Scientific evaluation results. The mean deviation from input values and associated standard deviation are reported for the main morphological parameters. Specifically, they are: the centre position (x and y), the integrated magnitude, effective radius (${\rm R_{e}}$), S\'ersic index (n), axis ratio (b/a) and position angle (PA) of the galaxy.}
\label{tab:nscier}
\begin{tabular}{lrrrr}
\hline\hline
Parameter          & OBS. [CONTROL]     & REC.               & NONE             \\
\hline
x[mas]             & $ 0.000 \pm 0.164$ & $-0.194 \pm 0.179$ & $-0.120 \pm 0.284$ \\
y[mas]             & $-0.045 \pm 0.105$ & $-0.030 \pm 0.105$ & $ 0.000 \pm 0.149$ \\
int. mag. [mag]    & $-0.002 \pm 0.019$ & $-0.009 \pm 0.016$ & $ 0.119 \pm 0.017$ \\
${\rm R_{e}}$[mas] & $-0.807 \pm 3.005$ & $ 2.497 \pm 2.272$ & $ 69.5773 \pm 6.055$ \\
n                  & $-0.105 \pm 0.324$ & $-0.841 \pm 0.188$ & $-1.640 \pm 0.087$ \\
b/a                & $ 0.002 \pm 0.005$ & $ 0.042 \pm 0.005$ & $ 0.217 \pm 0.005$ \\
PA[deg]            & $ 0.095 \pm 0.167$ & $-0.621 \pm 0.150$ & $-0.737 \pm 0.229$ \\
\hline
\end{tabular}
\end{table*}
\begin{figure}[t]
  \centering
     \includegraphics[height=.7\textheight]{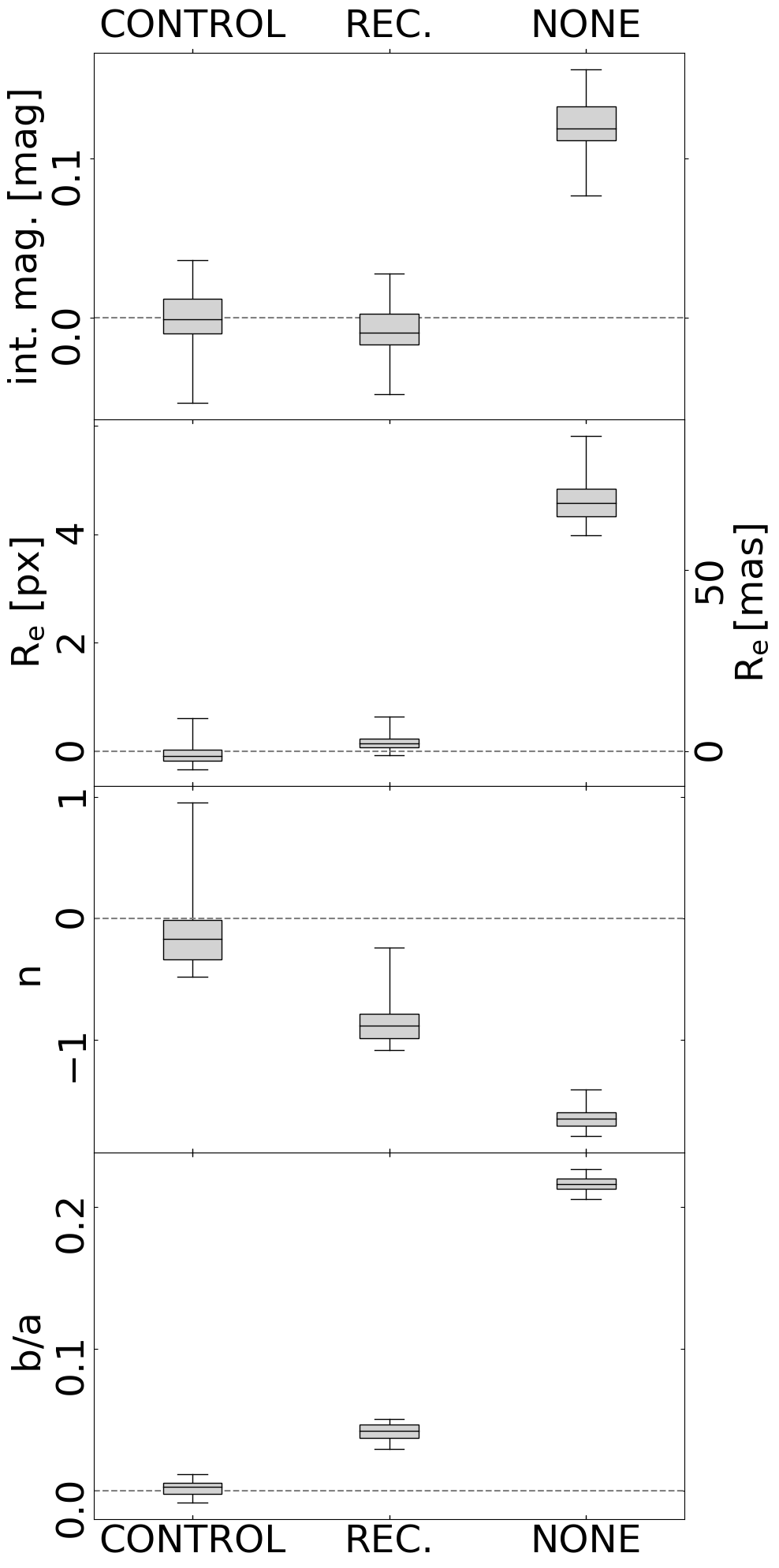} 
     \caption{Box plots of the measured deviations from input values for integrated magnitude, effective radius ${\rm R_{e}}$, S\'ersic index ${\rm n}$, and axis ratio $b/a$ from top to bottom. The control measure refers to the use of the observed PSF (the one that was used for simulating the observations). The actual measure (REC.) has been obtained using the reconstructed PSF as the GALFIT input. For comparison, the box plots are shown also for the case when no PSF at all was provided as input to the software.}
     \label{fig:sedevbp}
\end{figure}

With the aim of testing the applicability of the reconstruction to an ideal science case, we defined a general method to translate the differences between observed and reconstructed PSF into errors on the scientific measurements. In general, this would ultimately allow the definition of constraints on the reconstructed PSFs in order to match the requirements for the precision on selected science measurements. Limiting as much as possible the effects of observation-speicifc factors, we envisaged a representative case where the reconstruction of the PSF is critical: the case of a distant (e.g. $z \sim 1.5$) galaxy modelled using a S\'ersic profile, without adding any substructures. The simulated galaxy have effective radius comparable to the FWHM of the PSF, in order to make fundamental the knowledge of the obesrved PSF for its photometric and morphological characterization. Similarly, we  chose to simulate a bright galaxy to assess the effects of the outer, faint regions of the PSF.
In particular, we used the nighttime reconstructed PSF to investigate the accuracy achieved when modelling the 2D light profile of the galaxy simulating it with GALFIT\cite{2010AJ....139.2097P}. 
The parameters of the  simulated  galaxy are: integrated magnitude ${\rm m_{Vega}}=17.5\,{\rm mag}$; S\'ersic index ${\rm n}=3.9$; effective radius $R_e=94\,{\rm mas}$ ($6\,{\rm px}$, corresponding to $\sim2\,{\rm FWHM}$); axis ratio of $0.28$. These values are typical for a red nugget (an early-type galaxy) at redshift $z\sim1.5$, with the exception of the integrated magnitude\cite{2009ApJ...695..101D} .
For generating the simulated SOUL+LUCI observations, we make extensive use of SimCADO \cite{2016SPIE.9911E..24L,2020SPIE11452E..1ZL} , specifically tuning it for the purpose. The simulated observations are generated convolving the S\'ersic model with the nighttime observed PSF. 
We fix the total exposure time at $54000\,{\rm s}$, simulating $180$ frames of $300\,{\rm s}$ each.
To ensure a statistical significance of the results, we generated $100$ mock observations of the same model galaxy, adding at each time random white noise. A zoom-in of a single mock observation is shown in the left panel of Figure \ref{fig:nscier}.
The measurements have been performed on each mock observation with GALFIT giving different PSFs as input to the software: the observed PSF (the control measure), the reconstructed PSF (the actual measure) and no PSF at all (for comparison). The best-fit GALFIT model made with the reconstructed PSF is shown in the central panel of Figure~\ref{fig:nscier}. The match between the simulated galaxy and the model can be qualitatively estimated obesrving the normalized residuals, presented in the right panel of Figure~\ref{fig:nscier}. No evident residual structures can be detected.

Quantitatively, all the morphological parameters are in good agreement. We report in Table~\ref{tab:nscier} the mean differences between recovered and input values along with the associated standard deviations. The box plots, relative to the main parameters only, are shown in Figure~\ref{fig:sedevbp}. They represent the distribution of the deviations from input values for the $100$ simulations. The loss of accuracy when no PSF are used for the measure is noticeable, confirming the need of information about the observed PSF in this particular case. 
From Table~\ref{tab:nscier} and Figure~\ref{fig:sedevbp} it can be also noted that the best-fit model obtained with the reconstructed PSF, has a $n$ value $2\sigma$ lower than expected.
As a general consideration, although an optimization of the analysis is possible, our two-dimensional fit still performed successfully and the galaxy can be correctly classified as an early type one having $2.5<{\rm n}<4$.

\section{CONCLUSION}\label{sec:fin}
The PSF-R technique allows the estimation of the PSF of an astronomical observation from synchronous WFS telemetry data, without any access to the focal plane data. This is an unavoidable solution when there is no bright and isolated star in the scientific field of view, suitable for deriving the
observed PSF. 
In the context of MICADO, a first light instrument of ELT, a PSF-R software is being developed\cite{2018JATIS...4d9003W}. The current status of the MICADO PSF-R service\cite{2020SPIE11448E..37S} development is discussed in Ref~\citenum{grazian_psfrstatus}. Here we report on the testing of its performance on real SCAO observations taken with SOUL+LUCI at LBT. An interesting aspect of the present work is that we successfully applied the PSF-R technique to real AO data acquired with a pyramid WFS. 
In particular, we were able to reproduce the observed PSF FWHM, EE, and SR with an accuracy of $2-4\%$. In addition, we have proposed a method to directly assess the impact of the differences between observed and reconstructed PSFs on scientific measurements. The focus of this first scientific evaluation of the reconstructed PSFs is the photometric and morphological characterization of an idealized compact early-type galaxy. The analysis suggests that all the scientific parameters are recovered with sufficient precision.

For what concerns the future development, two main paths will be followed: i) the analysis of a wider dataset of SOUL+LUCI data is ongoing, the goal is to test the PSF-R algorithm on a wider parameter space (see Ref.~\citenum{arcidiacono_lucipsfr}); ii) the extension of the PSF-R method to the off-axis case (see Ref.~\citenum{wagner_offaxispsfr}).

\acknowledgments 
This work has been partly supported by INAF through the Math, ASTronomy and Research (MAST\&R), a working group for mathematical methods for high-resolution imaging. Based on observations made at the Large Binocular Telescope (LBT) at Mt. Graham (Arizona,USA). The LBT is an international collaboration among institutions in the United States, Italy and Germany. LBT Corporation partners are: The University of Arizona on behalf of the Arizona university system; Istituto Nazionale di Astrofisica, Italy; LBT Beteiligungsgesellschaft, Germany, representing the Max-Planck Society, the Astrophysical Institute Potsdam, and Heidelberg University; The Ohio State University, and The Research Corporation, on behalf of The University of Notre Dame, University of Minnesota and University of Virginia.

\bibliography{lbtpsfr} 
\bibliographystyle{spiebib} 

\end{document}